\documentstyle[aps,twocolumn]{revtex}
\begin{document}
\title{Full observation of single-atom dynamics in cavity QED}
\author{H. Mabuchi, J. Ye, and H. J. Kimble}
\address{Norman Bridge Laboratory of Physics 12-33, California
Institute of Technology, Pasadena, CA 91125 U.S.A.}
\date{May 24, 1998}
\maketitle
\begin{abstract}
We report the use of broadband heterodyne spectroscopy to perform continuous measurement of
the interaction energy $E_{\rm int}$ between one atom and a high-finesse optical cavity, during
individual transit events of $\sim 250$ $\mu$s duration.  We achieve a fractional sensitivity
$\simeq 4\times 10^{-4}/\sqrt{\rm Hz}$ to variations in $E_{\rm int}/\hbar$ within a measurement
bandwidth that covers 2.5 decades of frequency (1--300 kHz).  Our basic procedure is to drop
cold Cesium atoms into the cavity from a magnetooptic trap while monitoring the cavity's
complex optical susceptibility with a weak probe laser.  The instantaneous value of the
atom-cavity interaction energy, which in turn determines the coupled system's optical
susceptibility, depends on both the atomic position and (Zeeman) internal state.  Measurements
over a wide range of atom-cavity detunings reveal the transition from resonant to dispersive
coupling, via the transfer of atom-induced signals from the amplitude to the phase of light
transmitted through the cavity.  By suppressing all sources of excess technical noise, we
approach a measurement regime in which the broadband photocurrent may be interpreted as a
classical record of {\em conditional} quantum evolution in the sense of recently-developed
quantum trajectory theories.
\end{abstract}

\noindent
Optical cavity quantum electrodynamics (QED) in the strong coupling regime \cite{Kimb94}
provides a unique experimental paradigm for {\em real-time} observation of quantum dynamical
processes at the {\em single-atom} level.  While spectacular advances have certainly been made
in the preparation and tomography of quantum states of motion for a single trapped ion
\cite{Meek96,Leib96}, all such experiments have involved the accumulation of ensemble-averaged
data over many successive realizations of the process being studied.  Recent studies of
single-molecule dynamics have likewise demonstrated the ``immediate'' detection of photochemical
\cite{Moer94} or conformational \cite{Stip98} events, but such experiments presently lack the
potential that cavity QED provides for observing quantum processes on a timescale that makes
coherent control/intervention a tangible possibility.  We wish here to look beyond the mere
detection of quantum jumps, and to focus on the development of a broadband, {\em single-shot}
measurement technique that achieves signal-to-noise ratio $> 1$ over a bandwidth that includes
all characterstic frequencies of a quantum dynamical process.

Real-time observation of quantum dynamics in {\em many-atom} systems has recently become an
important theme in atomic physics, with notable demonstrations involving vibrational
excitations of a trapped Bose-Einstein condensate \cite{Andr97} and the decay of coherent
oscillations of an ensemble of atoms in an optical lattice \cite{Kozu98,Guib96}.  In contrast
to programs like these, for which the scientific emphasis lies on noninvasive observation of
a system's intrinsic dynamical processes, experiments in single-atom cavity QED hold great
potential for enabling precise investigations of how measurement backaction {\em alters} the
dynamical behavior of a continuously-observed open quantum system
\cite{Cave87,Wise96,Herk96,Dohe98,Mabu98a}.

A sophisticated theoretical basis for understanding such issues is presently maturing in the
form of quantum trajectory theories \cite{Carm93,Wise93a,Wise93b,Zoll97}, but significant
technical challenges remain to be solved before definitive experiments can be performed in the
lab.  Our purpose in the present work is to report substantial progress towards surmounting
such obstacles in the context of cavity QED, and hence towards achieving the essential
experimental capabilities required to perform quantitative tests of measurement-based
Stochastic Master Equations.  We ultimately hope to be able to implement some recently-proposed
``applications'' of the continuous observation of dissipative quantum dynamics, in fields such
as quantum measurement \cite{Wise95,Mabu96c}, quantum chaos \cite{Alic96,Scha96}, and quantum
feedback control \cite{Wise94,Mabu96c,Dunn97,Wong97}.

This article focuses on a detailed description of our recent experiments that record the
complete time-evolution of interaction energy between one atom and a high-finesse optical
cavity, during individual transit events of $\sim 250$ $\mu$s duration.  With characteristic
atom-cavity interaction energies $E_{\rm int}/\hbar \sim 10$ MHz, we achieve measurement
sensitivities $S_g\simeq 4.5$ kHz$/\sqrt{\rm Hz}$ over a bandwidth that covers the dominant
rates of variation in $E_{\rm int}$ (1--300 kHz).  Unlike typical pump-probe measurements
of scattering dynamics in real ({\it e.g.} diatomic) molecular systems \cite{Zewa96,Gens98},
our experiments on the Jaynes-Cummings ``molecule'' yield a continuous time-domain record of
the atom-cavity coupling during each individual ``scattering'' event (transit).  The data
clearly illustrate variations caused by atomic motion through the spatial structure of the
cavity eigenmode and/or optical pumping among the atomic internal (Zeeman) states.  In certain
parameter regimes of the detuning and probe power, distinctive indications of the
quantum-mechanical nature of the atom-cavity coupling can be seen in the photocurrent recorded
from just a {\em single} atomic transit.  For large ($\ge 50$ MHz) atom-cavity detunings we
obtain phase-contrast signals induced by individual atomic transits, corresponding to a regime
of strong but dispersive coupling.  Phase-quadrature measurements of atomic motion have been
widely discussed in the quantum optics literature
\cite{Mart92,Stor94,Gerr96,Herk96,Wong97,Dohe98}, but the present work provides the first
experimental demonstration at the single-atom level.

Because of the standing-wave spatial structure of the cavity eigenmode, and the corresponding
rapid varation of the atom-cavity coupling strength over sub-wavelength distances, our data
should in principle display a {\em sensitivity} of $1.5\times 10^{-10}$ m/$\sqrt{\rm Hz}$ to
atomic displacements along the cavity axis.  Unfortunately we cannot realize this figure as a
{\em precision} for monitoring the atomic position, as we do not presently have any means of
separating signal variations due to motion through the standing wave from ``background''
contributions due to transverse motion or optical pumping.  In our concluding section, we
shall briefly discuss our motivations for further work to disambiguate the nature of rapid
variations in our data.

\section{Basic theoretical description}

In simple terms, our experimental procedure is to drop a cloud of cold Cesium atoms from a 
magnetooptic trap (MOT) into a high-finesse optical cavity, while continuously monitoring the
cavity's complex susceptibility with a weak probe laser \cite{Mabu96b,Hood98}.  By limiting the
number of atoms in the initial cloud, we can easily reach an operating regime in which atoms
transit the cavity only one at a time.  Using broadband heterodyne detection and a high-speed
digitizer, we continuously record both the amplitude and phase of the transmitted probe beam
during $\sim 50$ ms time windows.  Each window typically contains from zero to five atom
transit signals.

The elementary theoretical description of such a measurement employs steady-state solutions
of the nonselective Master Equation for a {\em stationary} two-level atom coupled to a single
electromagnetic mode via the Jaynes-Cummings interaction Hamiltonian.  Atomic center-of-mass
motion and optical pumping among Zeeman states can only be included in this treatment by
allowing for a time-dependent atom-cavity coupling strength.  Although this type of approach
cannot make predictions about dynamical variations in the coupling strength, it does provides
a quantitative basis for interpreting some time-independent features of our data.  Our use of
this ``adiabatic'' model may be justified to a certain extent by the separation of timescales
that we achieve in optical cavity QED with laser-cooled atoms.  In optimal cases, the
atom-cavity coupling strength should vary by as little as a factor of $10^{-4}$ over the system
damping time $\sim 30$ ns.

If we let $\rho$ denote the density operator for the joint state of the atom and cavity, the
nonselective Master equation (in the electric dipole and rotating-wave approximations) reads:
\begin{eqnarray}
{\dot\rho}&=&{-i\over\hbar}\left[{\hat H}_0,\rho\right] + \gamma_\perp\left(2{\hat\sigma}\rho
{\hat\sigma}^\dagger - {\hat\sigma}^\dagger{\hat\sigma}\rho - \rho{\hat\sigma}^\dagger
{\hat\sigma}\right)\nonumber\\
&&+\left(\kappa_a+\kappa_b+\kappa_c\right)\left(2{\hat a}\rho{\hat a}^\dagger - {\hat a}^\dagger
{\hat a}\rho - {\hat a}^\dagger{\hat a}\rho\right),\label{eq:ME}\\
{\hat H}_0&=&\hbar\Delta{\hat a}^\dagger{\hat a} +
\hbar\Theta{\hat\sigma}^\dagger{\hat\sigma} + \sqrt{2\kappa_a}{\cal E}\left({\hat a}+
{\hat a^\dagger}\right) + H_{\rm int},\\
H_{\rm int}&=&\hbar g_0 e^{-\left(y^2+z^2\right)/w^2}
\cos\left(k_Lx\right)\left[{\hat a}{\hat\sigma}^\dagger+{\hat a}^\dagger
{\hat\sigma}\right].
\end{eqnarray}
Here $\gamma_\perp$ is the atomic dipole decay rate, $\kappa_a$ is the cavity field decay rate
through the input mirror (through which the probe laser is injected), $\kappa_b$ is the cavity
field decay rate through the output mirror, $\kappa_c$ is the cavity field decay rate due to
intracavity scattering/absorption losses,  $\Delta\equiv \nu_a-\nu_p$ is the atom-probe
detuning, $\Theta\equiv \nu_c-\nu_p$ is the cavity-probe detuning, and the coupling strength
$g_0$ is equal to half the maximum single-photon Rabi frequency.  We treat the atomic
center-of-mass coordinates $x,y,z$ as c-number parameters, with the $x$-axis conciding with the
cavity axis and $z$ parallel to gravity.  The Gaussian waist of our cavity mode is $w\simeq45$ 
$\mu$m.  Note that we have written the Master Equation in a frame rotating at the drive
frequency, so ${\cal E}$ is a constant term proportional to the complex amplitude of the
driving field.

To find the steady-state density operator as a function of driving strength and various
detunings, we simply set ${\dot\rho_{ss}}=0$ and solve for $\rho_{ss}$ using linear algebra.
The expected amplitude and phase of the heterodyne photocurrent may then be computed as
\cite{Coll87,Wise93a}
\begin{equation}
\langle i_{het}\left(t\right)\rangle=\eta f_{L}^{1/2}\sqrt{2\kappa_b}\exp\left[i\left(
\Omega_Lt+\phi_L\right)\right]{\rm Tr}\left[\rho_{ss}{\hat a}\right],
\label{eq:hetphot}
\end{equation}
where $\eta$ represents the overall photodetection efficiency (including propagation losses
between the cavity and photodetectors, heterodyne efficiency, and detector quantum efficiency),
$f_L$ and $\phi_L$ represent the photon flux and phase of the (optical) local oscillator, and
$\Omega_L$ is the frequency of the optical local oscillator relative to the rotating frame
(cavity driving field).  In the experiment, we mix $i_{het}\left(t\right)$ with an rf local
oscillator at the frequency $\Omega_L$ (which ranges between 40 and 190 MHz) and separately
record the in-phase and quadrature components of the slowly-varying envelope (with an analog
bandwidth of 300 kHz).

Note that the presence of an intracavity atom can influence the heterodyne photocurrent only
via the interaction Hamiltonian $H_{\rm int}\propto \hbar g({\bf r})$, where
\begin{equation}
g({\bf r})\equiv g_0e^{-\left(y^2+z^2\right)/w^2}\cos\left(k_Lx\right).
\end{equation}
In a two-level approximation for the atomic internal dynamics, and for a classical treatment
of the atomic center-of-mass degrees of freedom, {\em all} steady-state properties of the
atom-cavity system are strictly determined by the value of $g({\bf r})$ once the parameters
($\kappa_a$,$\kappa_b$,$\gamma_\perp$,$\Delta$,$\Theta$,${\cal E}$) have been specified.
This includes the quantity $\langle H_{\rm int}\rangle$, which represents one possible measure
of the ``interaction energy'' between atom and cavity.  Given that $g({\bf r})$ appears to be a
more fundamental measure of the interaction strength, however, we have adopted the convention
$E_{\rm int}\equiv\hbar g({\bf r})$.

The atom-cavity interaction can be treated semiclassically using the optical bistability
state equation (OBSE) \cite{Lugi84}.  The OBSE is traditionally written in terms of the scaled
field variables $x$ and $y$ (not to be confused with the atomic coordinates), with the
correspondence
\begin{equation}
x\equiv {\langle{\hat a}\rangle\over\sqrt{m_0}},\quad y\equiv {\sqrt{2\kappa_a}{\cal E}\over
\left(\kappa_a + \kappa_b\right)\sqrt{m_0}},
\end{equation}
where the saturation photon number $m_0$ is given by ${\gamma_\perp^2\over 2g^2}$.  For a given
driving strength ${\cal E}$, the expected intracavity field amplitude can be found by inverting
the equation
\begin{equation}
y=x\left(1 + {2C\over 1 + \delta^2 + x^2} + i{\phi-2C\delta\over 1 + \delta^2 + x^2}\right),
\label{eq:OBSE}
\end{equation}
where we again work in a rotating frame at the drive frequency, $\delta\equiv \left(\omega_a
-\omega_p\right)/\gamma_\perp$ represents the scaled atom-probe detuning, $\phi\equiv \left(
\omega_c-\omega_p\right)/\left(\kappa_a+\kappa_b+\kappa_c\right)$ represents the scaled
cavity-probe detuning, and the ``cooperativity'' $C$ is defined by
\begin{equation}
C\equiv {g^2\over 2\left(\kappa_a + \kappa_b + \kappa_c \right)\gamma_\perp}.
\end{equation}
Note that the dependence of $g$ (and thereby $C$ and $m_0$) on the atomic position and internal
state is implicit.  A semiclassical prediction for the heterodyne photocurrent is obtained by
substituting $x\sqrt{m_0}$ for ${\rm Tr}\left[\rho_{ss}{\hat a}\right]$ in equation
(\ref{eq:hetphot}).  

\section{Experimental apparatus and procedures}

Figure \ref{fig:apparat} provides a general overview of the apparatus, indicating the schematic
arrangement of various components to be described below.   The diode laser setup for forming
the Cs MOT is not shown.

\subsection{High-finesse microcavity}

We use a Fabry-Perot high-finesse microcavity (``physics cavity'') consisting of two spherical
mirrors with 1 m radius of curvature \cite{REO}.  The cavity was constructed with a mean length
$l\simeq 107.5$ $\mu$m, which we inferred from the cavity's measured free spectral range of
$1.395\times 10^{12}$ Hz.  The measured $l$ and specified radii of curvature geometrically
determine the cavity's electromagnetic mode volume for TEM$_{00}$ modes near 852 nm
\cite{Sieg86}.  Together with the dipole decay rate $\gamma_\perp/2\pi \simeq 2.6$ MHz
for the Cs 6P$_{3/2}$ level \cite{Tann93}, this determines our optimal coupling constant
$g_0/2\pi$ to be $\simeq 11$ MHz for $\sigma_\pm$ transistions (specifically the $6S_{1/2}\left(
F=4,m_F=\pm 4\right)\rightarrow 6P_{3/2}\left(F=5,m_F=\pm 5\right)$) and $\simeq 6$ MHz for
$\pi$ transitions ($6S_{1/2}\left(F=4,m_F=0\right)\rightarrow 6P_{3/2}\left(F=5,m_F=0\right)$)
within the D2 Zeeman manifold \cite{Kimb94,Schm94}.

In order to allow cold atoms to fall into such a short cavity, we found it necessary to have
the mirror manufacturer reduce the substrate diameters from the standard value of 7.75 mm down
to 3 mm \cite{REO}.  This reduced the ``sagittal depth'' of the curved mirror substrates and
allowed us to maintain a gap of $\sim 100$ $\mu$m around the edge of the cavity.  Machining of
the mirror substrates was performed {\em after} they had been superpolished and coated, but this
process did not seem to degrade the mirror reflectivities significantly.  The nominal combined
transmission and loss per mirror, before machining, was $1.5\times 10^{-5}$.  Direct
measurements of the cavity finesse yield ${\cal F}\simeq 217,000$ at an optical wavelength of
852.36 nm, consistent with a combined mirror transmission and loss $T^*\simeq 1.45\times
10^{-5}$.  This value of $T^*$ is inferred from the measured value of $l$ and the measured
cavity HWHM $\kappa/2\pi\simeq 3.21$ MHz.

The cavity used throughout the work described in this paper suffers from a rather pronounced
birefringence, which for TEM$_{00}$ modes near 852 nm induced a splitting of $8\pm 2$ MHz
between linearly-polarized eigenmodes.  It is not entirely clear whether this birefrigence
is a result of the substrate machining for diameter-reduction, a property of the coatings, or
something associated with the mirror-mounting procedure used for this particular cavity.  We
note that recent efforts by other members of our group \cite{Hood98} have produced a cavity of
approximately the same finesse with greatly-reduced birefringence (by a factor $\sim 25$),
using mirrors from a different coating run and with great care taken to minimize cavity
misalignments and stress on the mirror substrates.

Our two mirror substrates are mounted in vee-grooves atop independent aluminum blocks, with a
piezoelectric actuator between the blocks for active servo-control of the mirror separation
(cavity length).  The cavity mount sits on a stack of alternating OFHC copper blocks and viton
o-rings for passive vibration isolation, all within an ion-pumped vacuum chamber whose
background pressure was typically $\sim 10^{-8}$ torr (inferred from the ion pump current).  In
order to bring the MOT as close as possible to the central axis of the physics cavity, we had
to use a rather open (and non-magnetic) mount design, leading to some compromises in the way of
mechanical stability.  Sitting on the vibration-isolation stack and under vacuum, we found that
the native noise spectrum of the cavity length extended out to about 4 kHz, with one prominent
resonance at 50 Hz (which we attribute to a transmission resonance of the isolation stack).
Due to a set of PZT-actuator resonances above 10 kHz, we have ultimately been limited to a
unity-gain bandwidth $\simeq 1$ kHz for the cavity stabilization servo.

\subsection{Laser and cavity locking schemes}

Although the principle aim for this experiment is to stabilize the cavity length at some
precise offset $\Delta/2\pi\sim 0-100$ MHz from the Cs D2 resonance at 852.359 nm, the strong
atom-cavity coupling places severe restrictions on the optical power that can be used for the
purpose of generating an error signal.  On resonance, the saturation intracavity photon
number for our cavity is as small as $m_0\equiv \gamma_\perp^2/2g_0^2\simeq 0.1,$ which sets a
fiducial cavity throughput of $4\pi\kappa_b m_0\sim 1$ pW.  With such low optical power it
would be extremely difficult to obtain a high-quality error signal for locking the physics
cavity.  Other experiments performed in our group have circumvented this problem by using a
chopped locking scheme, in which a strong ``lock beam'' alternates with a weak ``probe beam''
at 50\% duty cycle and $\sim 1-4$ kHz frequency \cite{Turc95a,Mabu96b,Hood98}.  However, such a
strategy inherently limits the servo unity-gain bandwidths to $\sim 100$ Hz at best, and would
not be appropriate for future experiments with atoms trapped inside the cavity for long periods
of time.

In this work we have developed an alternative locking scheme for the physics cavity, which
employs an auxilliary diode laser at 836 nm to monitor the cavity length on a different
longitudinal mode than that which couples strongly to intracavity Cs atoms.  We use a commercial
grating-stabilized diode produced by New Focus (Santa Clara, California).  With a 16 nm
detuning, we can send $\sim 40$ nW through the cavity and incur an AC Stark shift of only
$\sim 60$ kHz for the atomic resonance at $\nu_a$.  Using an EG\&G
avalanche-photodiode/transimpedance amplifier module (model C30998) for AC detection of
transmitted 836 nm light, we obtain an FM error signal (modulation frequency 3.8 MHz) with
signal-to-noise ratio (SNR) $\sim 20$ at 30 kHz bandwidth.

The 836 nm diode laser is stabilized to an auxilliary ``transfer cavity,'' which consists of a
pair of 25 cm radius-of-curvature mirrors at $\simeq 16$ cm separation.  One of the mirrors is
mounted on a piezoelectric actuator to allow cancellation of DC drift and low-frequency noise.
The transfer cavity has a linewidth $\simeq 100$ kHz at both 836 nm and 852 nm, with an overall
mode spacing $\sim 300$ MHz.  The transfer cavity is also used for pre-stabilization of the
Ti:Sapphire laser, and some of the Ti:Sapphire light is used in a Cs modulation-transfer
spectrometer \cite{Shir82} to provide an absolute reference for the transfer cavity length.
From run to run we used one or two acoustooptic modulators to offset the
Ti:Sapphire/transfer-cavity lock point by +140, +87, or +43 MHz relative to the Cs
6S$_{1/2}(F=4)\rightarrow$ 6P$_{3/2}(F'=5)$ transition.

By comparison of the Ti:Sapphire and diode laser error signals in their respective locks to the
transfer cavity, we infer that the relative rms jitter between them is $\le 10$ kHz.  For both
laser locks we use the Pound-Drever-Hall technique \cite{Drev83} of detecting an FM signal in
reflection from the tranfser cavity.  The stability of the transfer cavity resonances with
respect to atomic Cs lines in a vapor cell was such that we did not see any relative jitter
beyond the measurement noise in our modulation-transfer spectrometer (SNR $\sim 50:1$ in
30 kHz bandwidth).

The Ti:Sapphire stabilization employs two feedback loops, one with $\sim 10$ kHz bandwidth to
the tweeter inside the ring laser and another with $\sim 100$ kHz bandwidth to a VCO-driven,
double-passed acoustooptic modulator (AOM) just outside the laser cavity \cite{Vass90,Boyd91}.
The mean frequency of the AOM is 76 MHz, and the error signal going to the VCO has a lower
cutoff of $\sim 10$ kHz to prevent DC drifts.  We note that use of the AOM is crucial for
achieving high stability of the Ti:Sapphire frequency.  The diode laser servo utilizes both
feedback to the grating PZT and direct modulation of the injection current, achieving an
overall unity-gain bandwidth $\sim 1$ MHz.

Having locked both the diode laser and Ti:Sapphire to modes of the transfer cavity, which
itself is locked to Cs, we use a travelling-wave electrooptic modulator to generate an rf
sideband of the diode laser at $f_0\sim 200-500$ MHz.  Either the upper or lower
sideband is used to derive an FM error signal for locking the physics cavity by dithering
$f_0$ at 3.8 MHz, thus allowing us to achieve arbitrary placement of the physics cavity
mode near 852 nm via the tunability of $f_0$.

\subsection{Evaluation of the physics-cavity servo}

Our basic requirement for the quality of the physics-cavity servo was that relative jitter of
the cavity resonance and the probe laser frequency should not contribute a significant amount
of noise in the heterodyne photocurrent.  Hence the relevant comparison to make is between the
noise in both quadratures of a demodulated beatnote and the photocurrent fluctuations produced
by the local oscillator alone.  At 200 kHz bandwidth and with a probe beam strength such that
$m\equiv\vert\langle{\hat a}\rangle\vert^2=1.5$ in the empty cavity, the standard deviations
of the phase and amplitude quadratures of the transmitted probe beam were measured to be 1.01
and 1.39 (respectively) relative to those of the local oscillator alone.  Note that we have
estimated the standard deviation of the quadrature-amplitude signals produced by our
optical local oscillator to be only a factor of 1.05 above the theoretical shot-noise limit
(see below).  We therefore believe that our overall excess noise factor $\beta\simeq 1.5$.

By taking some simultaneous recordings of the heterodyne photocurrent and the physics-cavity
error signal, we were able to verify directly that the atomic transits do not affect the
physics cavity servo.

\subsection{Cesium MOT}

To provide a source of cold Cs atoms, we used a standard magnetooptic trap loaded directly
from a thermal beam \cite{Cabl90}.  Our choice of thermal-beam loading, as opposed to loading
from a background vapor, was driven by an attempt to prevent accidental coating of the physics
cavity mirrors with Cs.  In more than two years of service, we did not detect any significant
($>5\%$) change in the cavity finesse.  Pre-cooling of the Cs beam was not necessary for this
experiment, as we required only a very low rate of delivering single cold atoms into the
cavity mode volume.  Our MOT employs a six-beam configuration, and we orient the anti-Helmholtz
coils for the trap so that their symmetry axis is parallel to that of the optical cavity.  This
leads to a MOT laser beam geometry with one beam axis running parallel to and just above the
cavity, plus two beam axes in the plane of the mirror surfaces (Figure \ref{fig:motconfig}).  

The light for the MOT was provided by a pair of grating-stabilized diode lasers (SDL 5421-G2),
one tuned to the Cs 6S$_{1/2}(F=4)\rightarrow$6P$_{3/2}(F'=5)$ cycling transition for trapping
and the other to 6S$_{1/2}(F=3)\rightarrow$6P$_{3/2}(F'=4)$ for repumping.  Each trapping beam
had $\sim 1$ cm diameter and anywhere from 40 $\mu$W to 4 mW of optical power, depending on how
many atoms we were trying to send into the physics cavity.  We typically used a Cs reservoir
temperature of 60-80 C for the thermal beam, which effused through a 200 $\mu$ pinhole and
travelled an overall distance of $\sim 60$ cm to the trapping region (with a cold mechanical
collimator in the way to reduce loading of the ion pump).  With an anti-Helmholtz field of
around 25 G/cm, we could load up to $\sim 2\times 10^5$ atoms into a millimeter-sized cloud,
whose mean temperature we estimate to be $\sim 100$ $\mu$K based on fluorescent imaging of free
expansion.  This temperature estimate is also supported by the spread in arrival times of
individual atoms falling into the cavity.

When running the experiment we would load the MOT for about 0.5 s, then drop it by quickly
turning off the trapping beams with an AOM (using an rf switch with $\sim 45$ dB attenuation).
After the trapping beams were thus extinguished, we would ramp down the anti-Helmholtz field
according to an RC-filtered step with $\sim 4$ ms time constant.  The repumping beam was left
on all the time, so that falling atoms would be shelved in the $F=4$ ground hyperfine level
before entering the cavity.  No specific preparation was performed with respect to the atomic
Zeeman states.  Dropping $\sim 2\times 10^5$ atoms we would generally see 30-50 atoms falling
through the central part of the cavity mode volume, so for single-atom transit data we had to
reduce the trapping beam power by a factor of $100$ to reach 0-2 atoms per drop.  The overall
repitition rate for the trap-drop cycle was typically 0.6 Hz.

\subsection{Probe generation and photodetection}

We used a balanced-heterodyne setup in order to achieve high-efficiency, zero-background
photodetection of $\sim 1$ pW levels of 852 nm light transmitted through the physics cavity.
The frequency difference between cavity probe light and the optical local oscillator for
heterodyne detection was between 40-190 MHz, depending upon our choice for the atom-probe
detuning.  The probe light was generated from the Ti:Sapphire output by cascading a +200 MHz
AOM and a tunable travelling-wave electrooptic modulator, which was driven between -245 MHz
and -440 MHz to produce the desired atom-probe detuning.  This indirect method was required to
prevent contamination of the heterodyne photocurrent by electronic noise at the heterodyne
frequency.

Light leaving the physics cavity first hit a color-separation mirror which reflected $> 99\%$
of the 852 nm light but transmited $\approx 30\%$ of the 836 nm light, allowing us to recover
an error signal for locking the physics cavity (see above) without compromising the overall
detection efficiency for the probe field.  Residual 836 nm light going to the heterodyne setup
amounted to only $\sim 30$ nW and had negligible effect on the photocurrent of interest.

The local oscillator (LO) for the optical heterodyne was spatially cleaned by a ${\cal F}\sim
1000$ Fabry-Perot cavity (linewidth $\sim 1$ MHz), which also served to strip off spectral
noise at 76 MHz associated with the AOM servo for stabilization of the Ti:Sapphire frequency.
The cleaning cavity was locked using the Pound-Drever-Hall method \cite{Drev83} with FM
sidebands at 24 MHz, which likewise had to be kept weak in order not to saturate the AC gain of
the heterodyne photodetectors.  We used a total of $\sim 2$ mW in the LO, which generated a
shot-noise level $\sim 5$ dB above the electronic noise of the photodetectors in the frequency
range of interest.

The difference photocurrent from the balanced heterodyne detectors was amplified up to
-50 dBm or higher, then divided by a $0^\circ$ rf splitter.  An independent signal generator
was used to produce an rf local oscillator at the heterodyne frequency, and it was halved using
a $90^\circ$ rf splitter.  The two identical copies of the photocurrent were mixed with the
in-phase and quadrature copies of the rf LO to produce an orthogonal pair of quadrature
amplitude (QA) signals at baseband.  The QA signals were further amplified, and passed through
300 kHz analog filters with a roll-off of 12 dB/octave.  We used a 12-bit ADC to sample both
QA's simultaneously at a rate of 10 MHz per channel, which is sufficiently high to avoid signal
aliasing completely.  Following each drop of the trap, we continuously recorded both QA's
for a data acquisition window of 50 ms and streamed the data to a hard drive for offline
processing following the experimental run.

Ideally, we would like the data acquisition procedure just described to yield directly the
amplitude and phase quadrature amplitudes of light transmitted through the cavity.  If we
write the transmitted optical field as ${\cal A}(t)e^{-i\omega_p t}$, where ${\cal A}(t)$ is
a slowly-varying complex amplitude, the amplitude $x_a$ and phase $x_p$ quadrature-amplitudes
are defined by ${\cal A} \equiv x_a + ix_p$.  With respect to the quantum-mechanical theory
of the Master Equation (\ref{eq:ME}), ${\cal A}\propto \langle {\hat a}\rangle$.  Note that we
define ${\cal A}$ to have zero phase when the cavity is empty, so that $x_p$ should have
zero mean when there are no intracavity atoms.

Given the way that we generate the probe beam, however, we have no way of generating a
phase-locked rf local oscillator to recover $x_a$ and $x_p$ directly.  The phase of the
heterodyne photocurrent differs from the phase that the light has just after it leaves the
cavity because of fluctuations in the relative optical path length travelled by the signal
beam and optical local oscilator in reaching the photodetectors.  So the two signals produced
by mixing the photocurrent with the shifted and unshifted copies of our rf local oscillator
correspond to an orthognal, but {\em rotated} pair of quadrature amplitudes $x_1$, $x_2$:
\begin{equation}
\left(\matrix{x_1\cr x_2}\right)=\left(\matrix{\cos\phi & -\sin\phi\cr \sin\phi & \cos\phi}
\right)\left(\matrix{x_a\cr x_p}\right).
\end{equation}
Luckily, the characteristic timescales for fluctuations in the phase $\phi$ are quite long
($\sim 10-100$ ms, corresponding to acoustic disturbances) compared to the 250 $\mu$s duration
of an individual atom-transit signal.  In processing the recorded data to produce the plots
discussed below, we have therefore used an ``adaptive'' definition of the amplitude and phase
quadrature amplitudes.  Within a window of 2 ms preceding the signal of interest, we estimate
the instantaneous value of $\phi$ by determining the rotation of $x_1$,$x_2$ that produces one
quadrature ${\tilde x}_p$ with zero mean and one quadrature ${\tilde x}_a$ with positive mean.
Then ${\tilde x}_p$ is operationally defined to be the phase quadrature photocurrent, and
${\tilde x}_a$ is the amplitude quadrature photocurrent.

Figure (\ref{fig:hmtlong}) shows an example of a 15 ms segment of our quadrature amplitude
data.  Note that some excess low-frequency noise can still be seen in the phase quadrature,
which constrains the lower end of our measurement bandwidth to $\sim 1$ kHz.  Six prominent
atom-transit signals, characterized by a sharp drop in the amplitude quadrature and a
simultaneous increase in the phase quadrature, can be seen between $t=0.009$ sec and $t=0.013$
sec.  In our subsequent discussions of the data, we shall focus on individual signal ``events''
of this type.

\section{Determination of intracavity photon number}

In order to make quantitative comparisons between data and theory, we need to calibrate the
strength of the driving field used in each experimental run.  As we will ultimately choose to
ignore DC optical phase offsets, the relevant quantity for us will be the number of photons $m$
that builds up when the cavity is empty.  This corresponds to $2\kappa_a\vert{\cal E}
\vert^2/\left(\kappa_a + \kappa_b + \kappa_c \right)^2$ in the Master Equation case and $\vert
y \vert^2/m_0$ for the OBSE.

Our strategy for determining the intracavity photon number during experimental runs has been
to work backwards from the heterodyne photocurrent observed with no atoms in the cavity, using
the expression \cite{Coll87}
\begin{equation}
\left({S^2\over N}\right)_{het} = 4T\eta\kappa_bm,
\label{eq:hetsn}
\end{equation}
where $S$ is the mean value of the demodulated amplitude quadrature signal (i.e. the
heterodyne photocurrent is of the form $S\cos(\omega t+\phi)$), $N$ is the mean-squared power
of fluctuations in the same signal due to (optical) shot noise, $T$ is the measurement
interval, and $m\equiv\vert\langle{\hat a}\rangle\vert^2$.  Note that the above
expression is valid for a coherent signal beam, which we assume to be the state of the light
transmitted through the cavity when no atoms are present.   For an accurate calibration, we
thus need to know the output mirror transmission $\kappa_b$, the total cavity loss (which can
be determined from measurements of the cavity finesse), and the overall photodetection
efficiency $\eta$.  In determining $m$ from the data, we typically chose $T$ on the order of
1 ms.  The following subsections provide further detail for each critical aspect of the
calibration.

\subsection{Evaluation of local oscillator noise}

Ideally, the quantity $N$ appearing in equation (\ref{eq:hetsn}) should coincide with the
photocurrent noise power $n$ observed when the signal beam is blocked.  This would allow us
to determine the intracavity photon number without having to calibrate the exact gains of the
photocurrent amplifiers, etc.  But the procedure is invalid if the optical local oscillator
carries excessive technical noise, so we have empirically checked the scaling of our nominal
$N$ with the DC optical power of the LO.  A fit of the data to the functional form $n = aP
+ bP^2$ yields $b/a\simeq 0.11$ mW$^{-1}$.  Given our typical operating LO power of 1 mW per
detector, and considering the relative magnitudes of other uncertainties, we approximate
$N\simeq n$.  Note that the LO does pass through an intensity stabilizer with $\sim 400$ Hz
bandwidth just before entering the cleaning cavity.

\subsection{Measurement of cavity decay rates}

We directly determined the total cavity field decay rate $\kappa_a + \kappa_b + \kappa_c$ by
making a calibrated measurement of the HWHM of a cavity TEM$_{00}$ resonance.  In order to
minimize systematic errors, we did this by using the heterodyne detectors to monitor the
transmitted optical power of a fixed-frequency Ti:Sapph probe beam while scanning the cavity
length.  The cavity length was always under servocontrol during the measurement, as we
generated the scan by stepping the frequency of the rf going to the travelling wave modulator
for the locking diode laser.  The total cavity field decay rate is then given by the resonance
HWHM measured in terms of the modulator rf scan, times a correction factor of the ratio of the
diode laser and Ti:Sapph wavelengths.  We find $\kappa_a + \kappa_b + \kappa_c\simeq 3.2$ MHz.

As the mirrors used to construct the cavity should be identical, we assign half the total
losses to each mirror.  Unfortunately, we did not manage to characterize the ratio of
intracavity losses $\kappa_c$ to transmission losses $\kappa_a + \kappa_b$ before we
accidentaly damaged the cavity.  Given that the mirror coatings we have previously received
from the same manufacturer have displayed very low scattering/absorption loss ($\sim 1.1\times
10^{-6}$) in the wavelength range of interest \cite{Remp92}, we have assumed $\kappa_b = 1.6$
MHz.

\subsection{Measurement of $\eta$}

Three principal factors determine the value of $\eta\equiv{\cal V}\varepsilon\tau$: the spatial
overlap $\sqrt{{\cal V}}$ between the signal beam and the optical local oscillator, the
photodetector quantum efficiency $\varepsilon$, and the (power) efficiency $\tau$ with which we
transfer light from the cavity ouput to the photodetectors.  Using an optical powermeter we
measured $\tau=0.9$, as well as $\varepsilon\simeq(0.68,0.71)$ for our two photodetectors.  We
measured ${\cal V}\simeq 0.65$ by producing a DC fringe between the signal beam and the optical
local oscillator, with both beams adjusted to a power level of 1 $\mu$W.  The power adjustments
were made upstream of the physics cavity and LO cleaning cavity, ensuring that no optical
misalignments were introduced in the process.  We obtained an independent measurement of
${\cal V}\varepsilon\simeq 0.4\pm 0.05$ from the heterodyne signal-to-noise ratio obtained with
a measured signal power of $10$ nW.  Note that the reasonable agreement between this latter
value and the directly-measured ${\cal V}\varepsilon$ provides further indication that our
optical local oscillator bears minimal excess noise, and that our assumptions
$\kappa_b=\kappa_a$, $\kappa_c\sim 0$ are valid.

Although it could have been avoided, we did suffer one additional loss of detection efficiency
due to gain and phase imbalance in the subtraction of heterodyne photocurrents.  Several weeks
after we took the data sets shown below, we realized that one photodetector output had an rf
signal 4 dB higher than the other (this figure includes both the deviation of our heterodyne
beamsplitter from being 50/50, the difference in photodiode quantum efficiencies, and the
difference in transimpedance gains), and a phase offset of 1 radian.  If we write
the two photocurrents (for a coherent signal beam) as $i_1 = {1\over\sqrt{2}}e^{-i\omega t} +
\xi_1$ and $i_2 = {-g\over\sqrt{2}}e^{-i\omega t + i\phi} + g\xi_2$ (where $\xi_1$ and $\xi_2$
are uncorrelated gaussian white noises with zero mean and unit variance), the difference
photocurrent is
\begin{equation}
i_- = {1 + ge^{i\phi}\over\sqrt{2}}e^{-i\omega t} + \sqrt{1 + g^2}\xi,
\end{equation}
where $\xi$ is again a gaussian white noise process with zero mean and unit variance.  The
complex imbalance $ge^{i\phi}$ should thus reduce the effective photodetection efficiency
(for the purpose of evaluating the heterodyne signal-to-noise ratio (\ref{eq:hetsn})) by
\begin{equation}
\eta\rightarrow {1\over 2}\eta{\vert 1 + ge^{i\phi}\vert^2\over 1 + g^2}.
\end{equation}
Using the measured $g\simeq 0.63$ and $\phi\simeq 0.85$ rad, $\eta\rightarrow 0.8\eta$, and
we quote an overall value of $\eta\simeq 0.32$.

\section{Numerical simulations}

In order to facilitate the interpretation of our data, we generated some rudimentary Monte
Carlo simulations of the heterodyne signals that we should see as atoms fall through the
cavity.\footnote{The computational work described in this section was conducted by H.M..}
Our code simulates three-dimensional, classical center-of-mass motion of individual atoms
under the influence of fluctuating forces due to their strong interactions with the cavity
field.  Gravity is also included in the kinematics, but plays only a minor role over the
$\sim 1$ ms duration of the simulations.

\subsection{Overall scheme for the simulations}

The basic scheme of the simulations is to precompute the values of the mean cavity-induced
force, the cavity-field-induced diffusion coefficient, and the steady-state (complex) amplitude
for the intracavity field as a function of atomic position.  Hence, {\em every} other degree
of freedom in the simulation is adiabatically eliminated and slaved to the atomic motion, which
is assumed to be the slowest and the ``stiffest'' process in the dynamics.  The mean force and
intracavity field are derived from steady-state solutions of the Master Equation (\ref{eq:ME}),
and the cavity diffusion is computed using the Quantum Regression Theorem \cite{Gard91}.  In
each timestep, the code first performs an interpolation on the precomputed tables of values to
determine the appropriate change in atomic momentum (which includes a stochastic increment
consistent with the local value of the diffusion constant), records an appropriate value for
the cavity output field, and then updates the atomic position and momentum.  The simplest
possible integration scheme is used (explicit Ito-Euler), yielding order $0.5$ convergence in
the timestep.  We chose a timestep of 7.5 ns in order to keep the run-times for the simulations
reasonable under Matlab on our Pentium II workstations, and this should have been sufficiently
small to keep the integration error below other sources of inaccuracy.

The principle shortcoming of this scheme is that it assumes the atomic velocity will remain
small enough that variations in the coupling strength $g$ will be negligible over timescales
on the order of $\kappa^{-1}$ and $\gamma^{-1}$.  If this condition is violated, then the
steady-state values of the mean cavity force and cavity output are no longer appropriate.
Another major approximation was made in treating only a two-level atom---optical pumping and/or
the associated opto-mechanical effects ({\it e.g.}, spontaneous forces) could certainly play an
important role in determining the shapes of our observed transit signals.  A third, less severe
approximation is that we have treated the stochastic process associated with recoil from
spontaneous-emissions as being statistically independent from the one associated with
dipole-force fluctuations.  The possibility that interesting effects could arise due to
quantization of the atomic motion is of course ignored as well.

Given the relative simplicity of our simulation scheme, it would be inappropriate to draw any
strong conclusions about our experimental data soley on the basis of apparent similarities with
the numerical results.  However, the simulations do provide an appealing {\em model} for the
atomic center-of-mass dynamics and associated heterodyne signals, with predictions that seem
to be fully consistent with what we see in the actual experiment.  This model suggests some
interesting interpretations for qualitative features of the data.  For example, the marked
asymmetry (in time) of most of our experimental transit signals may be associated with an
initial phase of atomic ``channeling'' in the dipole potential provided by the intracavity
optical field, followed by a sudden escape from local confinement due to cavity-field-induced
momentum diffusion.  Such a dynamical process, if it could be confirmed in some way, would bear
a strong resemblence to the Levy Walk behavior predicted in \cite{Mark96,DohePC} and observed
in \cite{Kato97}.  Our simulations also suggest that isolated dips and steps in the observed
transit signals may be associated with sudden changes in atomic localization relative to the
cavity standing wave.

Before presenting some results of the atomic-motion simulations, let us first describe the
numerical method used to compute the cavity diffusion coefficients.

\subsection{Computation of diffusion coefficients}

Following Doherty {\it et al} \cite{Dohe97}, we computed the diffusion coefficient associated
with dipole-force fluctuations according to
\begin{equation}
D=\lim_{t\rightarrow\infty}{\rm Re}\int_0^\infty\langle{\bf f}(t),{\bf f}(t-\tau)\rangle d\tau,
\label{eq:diffco}
\end{equation}
where ${\bf f}$ is the force operator
\begin{equation}
{\bf f} = -i\hbar \nabla g\left({\bf r}\right)\left({\hat a}^\dagger{\hat\sigma}_- -
{\hat a}{\hat\sigma}_+\right)
\end{equation}
and we define
\begin{equation}
g\left({\bf r}\right)\equiv g_0\cos(k_Lx)\exp\left[-\left(y^2+z^2\right)/w^2\right].
\label{eq:gofr}
\end{equation}
Note that the vector nature of the force operator comes only from the gradient $\nabla g\left(
{\bf r}\right)$.  Unlike the computation of mean forces and the expected intracavity field
amplitude, the evaluation of (\ref{eq:diffco}) requires an actual time-integration of the
cavity-QED Master Equation (\ref{eq:ME}).  Note that
\begin{equation}
\langle{\bf f}(t),{\bf f}(t-\tau)\rangle\equiv {\rm Tr}\left[{\bf f}e^{{\cal L}\tau}{\bf f}
\rho_{ss}\right] - \langle{\bf f}\rangle^2,
\label{eq:heisen}
\end{equation}
where ${\cal L}$ is the Liouville superoperator defined by
\begin{equation}
{d\over dt}\rho = {\cal L}\rho,
\end{equation}
with reference to the cavity-QED Master Equation (\ref{eq:ME}).  Knowing the steady-state
atom-cavity density matrix $\rho_{ss}$, one can evaluate the correlation function in
(\ref{eq:heisen}) by integrating the Master Equation for a time $\tau$ with ${\bf f}\rho_{ss}$
as an initial condition, multiplying the result by $\rho_{ss}$, and finally taking the trace.

The numerical integration was performed by a C++/MPI code on an SGI/Cray Origin-2000 cluster.
We used a truncated basis of 25 Fock states for the cavity mode, using an explicit Euler
integration with a 1 ps timestep and 5 $\mu$s total integration time.

Figure \ref{fig:cavdiff1} shows the results of such a calculation for the diffusion coefficient
as a function of atomic position along the standing wave, for an atom located on the cavity
axis.  Recall that the cavity mode function varies as $\cos(k_Lx)$ along the cavity axis, and
as a gaussian in the transverse dimensions.  Starting from the top, the four curves represent
$\Delta=0$, $10$, $30$, and $50$ MHz, all with a probe strength such that an average of 2
photons would build up in the cavity if it were empty and with $\Theta=0$.  Figure
\ref{fig:cavdiff2} shows similar results, all for $\Delta=50$ MHz, but with an average
intracavity photon number (for the empty cavity) of $m=8$ for the top curve, 4 for the middle,
and 2 for the lowest.  As the variation of $g$ along the standing wave is much more rapid than
in the transverse (Gaussian) directions, we approximated $\nabla g\left({\bf r} \right)\simeq
dg/dx$ in expression (\ref{eq:diffco}).  The computation of ${\bf f}$ for the mean force field
was of course three-dimensional.

Note that expression (\ref{eq:diffco}) gives the diffusion coefficient associated with dipole
force fluctuations only---it takes no account of recoils from spontaneous emission.  We
therefore incorporate a second diffusion process in the simulations, whose coefficient is
computed from the expectation value of the atomic excitation at each point in space:
\begin{equation}
D_{rec} = {\left(\hbar k_L\right)^2\over 25}\Gamma\langle{\hat\sigma}_+{\hat\sigma}_-\rangle.
\end{equation}
Here $\Gamma$ is the atomic spontaneous emission rate (Einstein A coefficient), and the number
$1/25$ comes from averaging over the angular distribution pattern for dipole radiation
\cite{Dohe97}.

\subsection{Discussion of results}

Figure \ref{fig:simex1} shows an example of the atomic position along the standing-wave $x$ and
the atomic velocity along the standing-wave $v_x$ from one typical simulation.  Also shown are
the mean value of the intracavity phase quadrature amplitude $\langle q_p\rangle\equiv
i\langle{\hat a}-{\hat a}^\dagger\rangle/2$, and a simulated heterodyne signal with
Gaussian noise added at the level appropriate to our experimental parameters (see equation
(\ref{eq:hetsn})).  Note that the quantity ($\langle q_p\rangle$ plus noise) is proportional to
the photocurrent ${\tilde x}_p$ that would be recorded in our experiment.  In the bottom row of
the figure, $\langle q_p\rangle$ and $\langle q_p\rangle$ plus shot-noise are shown at an
analog bandwidth 300 kHz in order to illustrate the effects of finite detection bandwidth on the
qualitative features of the data \cite{Dohe97}.  The results clearly suggests that the gross
features of the simulated signals are most strongly affected by atomic motion along the standing
wave, in that the overall gaussian profile associated with motion in the $z$-direction becomes
distorted by the ``envelope'' of the oscillatory variations due to motion along $x$ (recall
$g\left({\bf r}\right)=\cos(k_Lx)\exp\left[-\left(y^2+z^2\right)/w^2\right]$).

Looking at the results for $x(t)$ and $v_x(t)$, we see that the atom in this particular
simulation was mechanically confined within one well of the cavity standing wave until
$t\sim 810$ $\mu$s.  That is, $x$ did not vary by more than $\lambda/4$ and $v_x$ displays
the oscillations that one would expect to see for an atom trapped within a potential well.
At time $t\sim 810$ $\mu$s, however, we see that momentum diffusion finally pushes $v_x$ above
some threshold value such that the atom is able to escape from local confinement and ``fly
above'' the periodic dipole potential associated with the cavity standing wave.  Generally
speaking, one expects this type of escape to occur as the atom passes through the cavity axis
in the $-z$-direction and the standing-wave potential wells begin to decrease in depth.  The
expectation value $\langle q_p\rangle$ clearly reflects the qualitative features of the atomic
trajectory $x(t)$, although we also see that the details are lost in the measured signal due to
heterodyne shot-noise.  What survives in the simulated heterodyne signal is an overall
asymmetry in the envelope of the transit-signal, with a sudden ``step'' downwards at
$t\sim 800$ $\mu$s (as suggested by the dotted vertical line).

Figure \ref{fig:sims} shows four simulated signals with $\Delta=10$ MHz and an average
intracavity photon number of $m=2$, and two simulated signals with $\Delta=50$ MHz and $m=4$
($\Theta=0$ in all cases).  In each subplot, the upper trace shows the intracavity
amplitude quadrature $\langle q_a\rangle\equiv \langle{\hat a}+{\hat a}^\dagger\rangle/2$
versus time, and the lower trace (with zero mean value) shows the intracavity phase quadrature
$\langle q_a\rangle$ versus time.  Again, Gaussian noise has been added to the simulated
signals in order to reproduce the overall signal-to-noise ratio predicted by equation
(\ref{eq:hetsn}).  In all simulations, the atom was started at a position 7 gaussian waists
above the cavity with a vertical velocity of -47 cm/s.  The initial transverse position,
transverse velocity, position along the cavity standing-wave axis, and velocity along the
standing-wave were randomly assigned within parameter ranges that were narrow enough to ensure
that most simulated atom-drops produced a sizeable signal.

The simulated signals in Figure \ref{fig:sims} may be compared directly to the real data shown
in Figure \ref{fig:hmthbw}.  Like the simulation shown in Figure \ref{fig:simex1}, the signals
in Figure \ref{fig:sims} display a generic asymmetry and often contain an isolated dip or step.
The features of this type that are marked by an arrow near the horizontal axis are again
associated with sudden changes in the atomic localization relative to the cavity standing wave
(as determined by examining $x(t)$ from the simulations).

Although we have the benefit of knowing both the ``actual'' trajectory of an atom and the
corresponding heterodyne signal in our simulations, {\it a priori} knowledge of the atomic
trajectory is of course unavailable in our experiment.  Hence, the results of our numerical
simulations can only provide general guidelines for how we might try to infer something about
individual atomic trajectories from our heterodyne transit signals.  First of all, it appears
that signficant asymmetries in the observed signals could provide evidence that we do in fact
observe mechanical aspects of the atom-cavity coupling within individual atomic transits.
Relative to the approximations involved in our simulations however, one should bear in mind
that such asymmetries could also arise from optical pumping processes that may occur during
at atomic transit through the cavity.  Second, the simulations suggest that the transit signal
envelopes are most strongly influenced by atomic motion along the cavity standing wave, as
opposed to details of the motion along $y$ or $z$.  Third, isolated dips or steps in the signal
could be indicative of sudden changes in the atomic localization with respect to the standing
wave.  Again however, we should be careful to note that such features might also be caused by
intracavity optical pumping.

\section{Data}

Moving on to the experimental data, let us first discuss some transit signals displayed in the
same fashion as the simulations of Figure \ref{fig:sims}.  The quality of this data illustrates
our experimental ability to perform continuous, nearly quantum-noise limited measurement of the
atom-cavity interaction energy during individual scattering events.  In the second subsection
we shall display transit data on the complex plane, and present a quantitative comparison to
theoretical predictions of the atom-cavity system's complex optical susceptibility.  We shall
see that in optimal cases, sufficient information may be obtained from an {\em individual}
atom-transit signal to distinguish unambiguously between quantum and semi-classical models of
cavity QED.

Note that all of the data shown here were taken with cavity-probe detuning $\Theta\equiv
\nu_c-\nu_p=0$, while the atom-probe detuning $\Delta\equiv\nu_a-\nu_p$ and probe power
$m$ are varied.

\subsection{High bandwidth single-atom transits}

The basic unit of our experimental data is a continuous stream of quadrature-amplitude values
versus time, recorded for $\sim 50$ ms windows following each dropping of the MOT.  One
$\sim 15$ ms segment of such data is shown in Figure \ref{fig:hmtlong}.  This particular data
segment was taken with detunings $(\Delta=10,\Theta=0)$ MHz and with a probe power such that
an average of $1.5$ photons would build up in the cavity if it were empty ($m=1.5$).

Figure \ref{fig:hmthbw} shows six of the largest single-atom transit signals from our entire
data set, which covers detunings $\Delta$ from $-10$ to $+100$ MHz and probe powers $m$
from $1.5$ to $11$ photons.  The particular values of $\Delta$ and $m$ for each signal in
Figure \ref{fig:hmthbw} are displayed by the y-axis.  The data are shown at our full analog
bandwidth of 300 kHz and sampling rate of 10 MHz (12 bit resolution).  Note that we have
displaced the amplitude quadrature signals ${\tilde x}_a$ by $+400$ in order to prevent them
from overlapping with the phase quadrature signals ${\tilde x}_p$.

We have found that transit signals of maximal contrast tend not to have very much internal
structure, although there is a fairly prominant dip at the point indicated by an arrow in
subplot ``dat4,'' a set of three dips indicated by the arrow in subplot ``dat6,'' and
oscillatory structure in the signal of subplot ``dat5.''  The signal of subplot ``dat4'' is
really not so different in its overall structure than the simulations shown in Figure
\ref{fig:simex1} and in the ``traj10'' subplot of Figure \ref{fig:sims}.  One can see that the
shapes of the overall signal envelopes do vary substantially and are generally asymmetric in
time.  The distribution of the atom-transit signal between amplitude and phase quadratures
clearly depends upon the probe detuning, as will be discussed in greater detail below.  For the
data event recorded with $\Delta=0$, one sees that there is only a reduction in the power of
the transmitted probe without any shift in its phase.  At all other detunings, the atom-induced
reduction of amplitude quadrature is partly offset by an increase in the phase quadrature,
indicating a significant phase shift of the transmitted probe beam.  In subplot ``dat6'' taken
with $\Delta=50$ MHz, the signal is seen to primarly reside in the phase quadrature.

For signals such as those shown in Figure \ref{fig:hmthbw}, we may estimate a full-signal
to rms-noise ratio of approximately 2.5 for the phase quadrature and 4 for the amplitude
quadrature.  Combining the two signals, we have an overall ratio of 4.5.  The signal bandwidth
is 300 kHz, implying a relative sensitivity of $0.22/\sqrt{3\times 10^5}\simeq 4\times 10^{-4}$
Hz$^{-1/2}$.  Assuming that the largest signals correspond to atoms reaching the maximal
coupling strength of $g_0=11$ MHz, this sets our broadband sensitivity $S_g$ to time-varations
in $g\equiv E_{\rm int}/\hbar$ to be $S_g\simeq 4.5$ kHz/$\sqrt{\rm Hz}$.  Given our overall
photodetection efficiency $\eta\simeq 0.32$ and our independent assesment of the excess noise
factor $\beta\simeq 1.5$ (corresponding to technical noise on the optical local oscillator
power and on the physics cavity lock), we find that this value lies only a factor of $\beta/
\sqrt{\eta}\simeq 2.7$ above the fundamental quantum noise floor.  We maintain this high
sensitivity over 2.5 decades of signal frequency, from our full photodetection bandwidth of
300 kHz down to 1 kHz (at which point we are limited by residual technical noise in our
measurements of the phase quadrature).

Although a complete theoretical description of our experiment at the level of a Stochastic
Master Equation \cite{Wise93a} could in principle be formulated to include three-dimensional
quantized atomic center-of-mass motion, cavity birefringence, and the full manifold of internal
atomic Zeeman states, the resulting model would almost certainly be too complex to be useful
for quantitative interpretations of our existing data.  Nevertheless, we wish here to stress
that only a theory that explicitly treats the {\em quantum conditioning}
\cite{Carm93,Wise93a,Wise93b} of the atom-cavity evolution on the stochastic component of the
heterodyne photocurrent could provide a full account of the statistics and autocorrelation
properties of our atom-transit data.  In this crucial sense, we would claim that our recent
work may be distinguished from all other efforts in contemporary experimental physics.  Of
course, much remains to be done before we can offer definitive proof, so our continuing
efforts will largely be directed towards the unambiguous demonstration of quantum conditional
dynamics in cavity QED.

\subsection{Transit phasors}

In addition to displaying our atom-transit signals in the format of photocurrent versus time,
we can also construct parametric plots of ${\tilde x}_a$ versus ${\tilde x}_p$ to examine the
correlation induced between these two quantities by the atom-cavity interaction.  This is
equivalent to compiling a histogram of the complex amplitude of the optical field transmitted
through the cavity during an atomic transit, and hence to a continuous monitoring of the
atom-cavity system's complex optical susceptibility.  The underlying probability distributions
for such histograms are determined by the (evolving) Husimi Q-function for the intracavity
field \cite{Coll87,Wise93a}.

Many of our measurements have been conducted in a regime of strong cavity-driving fields ($m >
m_0g_0^2/\Delta^2$), with the consequence that significant effects of saturation and optical
nonlinearity can be seen in our experimental data.  As has been extensively discussed in the
cavity-QED literature, the nonlinear optical response of the atom-cavity system represents an
important experimental signature that may be used to distinguish between quantum and
semiclassical theoretical models for the atom-cavity interaction.  Prior to this work, four
groups \cite{Turc95b,Brun96,Thom98,Hood98} have reported nonlinear measurements in cavity QED.
In each of these previous experiments, the information gained per atomic transit about the
details of the system's nonlinear response was far less than in the data we shall present here.
Hence, we wish again to stress that the techniques developed in our most recent experimental
work have opened new possibilities for exploring quantum dynamics at the level of single
quantum realizations (trajectories), as opposed to the level of ensemble averages.

In Figures \ref{fig:hmtfig3} and \ref{fig:hmtfig2}, we display single-transit data on the
complex plane.  The experimental data are displayed as gray dots, with each dot representing
the values of ${\tilde x}_a$ and ${\tilde x}_p$ at some particular time during an atomic
transit.  In order to produce a set of dots from a transit signal such as those shown in
Figure \ref{fig:hmthbw}, the quadrature amplitude signals were first passed through an
anti-aliasing filter and then resampled every 10 $\mu$s to produce a discrete set of points
$\left\{{\tilde x}_a^i,{\tilde x}_p^i\right\}$.  For every $i$, a gray dot is placed at radius
$\sqrt{({\tilde x}_a^i)^2+({\tilde x}_p^i)^2}$ and polar angle $\tan^{-1}({\tilde x}_p^i
/{\tilde x}_a^i)$.

These transit phasors represent a simple way of looking at the correlation between the
amplitude and phase of light transmitted through the cavity, with time removed from the
picture.  We believe that this type of plot is the best format for comparisons to theory,
because atomic internal (optical pumping among Zeeman states) and external (motion through the
cavity eigenmode) dynamics should be factored out.  In the simplest approximation, these two
types of processes merely induce fluctuations in the atom-cavity coupling $g$, which should
only move the locus of our gray dots in and out along a curve parametrized by $g$.  The overall
shape of this curve should be dictated by the interaction Hamiltonian for the atom and cavity
mode, and is therefore quite easy to compute.

In Figure \ref{fig:hmtfig3}, we show a set of transit phasors taken at various probe detunings.
The data overlay theoretical curves predicted by the quantum and semiclassical theories.  The
quantum-mechanical predictions are computed by finding the steady-state solution of the Master
Equation (\ref{eq:ME}) for values of $g$ in the range $\left[0,g_0\right]$ and appropriate
values of $\Delta$ and $m$.  The solid curves represent an interpolation through the discrete
set of computed values for $\langle q_a\rangle$ and $\langle q_p\rangle$ as a function of $g$.
The semiclassical prediction is computed in the analogous manner, using the Optical Bistability
State Equation (\ref{eq:OBSE}) rather than the quantum Master Equation.  Note that both curves
must agree for $g=0$, and this common point of origin is marked in each subplot by a triangle.
The $g=g_0$ endpoint of the quantum curve is marked by a circle, and the $g=g_0$ endpoint of
the semiclassical curve is marked by an $\times$.  In the subplots of Figure \ref{fig:hmtfig3},
the grey dots represent an overlay of data from two individual atom-transit signals.  The two
particular signals displayed in each subplot were selected on the basis of having maximal
``contrast'' among all the signals from a given data set, under the assumption that the atoms
causing maximal signals should have come the closest to actually achieving $g=g_0$ on their way
through the cavity.  Quantum theory is seen to predict the observations quite well for all
the detunings shown in Figure \ref{fig:hmtfig3}.  

The quantum and semiclassical predictions shown in Figure \ref{fig:hmtfig3} do not differ
significantly except for the case of $\Delta=0$.  Generally speaking, the two theories are
known to agree in their predictions for weak-field response of the atom-cavity system but to
differ in their predictions regarding {\em saturation} of the optical response.  We here
present experimental evidence {\em from single atom transits} for the quantum character of
saturation in the atom-cavity system's response to near-resonant driving fields.

In Figure \ref{fig:hmtfig2} we show a set of transit phasors for fixed detuning $\Delta=10$ but
variable probe strength $m=2.8$, $4.4$, $7$, and $11$.  A clear discrepancy between the quantum
and semiclassical theory may be seen in this sequence of plots, with the data (gray dots)
showing significant preference for the quantum-mechanical predictions.  Each subplot in Figure
\ref{fig:hmtfig2} represents an overlay of data from three individual atom-transit signals,
in order to fill in the overall ``shape'' of the experimental transit phasors.  It should be
clear, however, that just a single data trace would suffice to determine that the experiment
matches much more closely with the quantum theory than the semiclassical theory.  We plan to
elaborate this result in separate publication \cite{Mabu98c}.

\section{Conclusions and future goals}

In summary, we have described the details of our recent experimental work to perform continuous
measurement of the interaction energy $E_{\rm int}\equiv \hbar g$ between one atom and an
optical cavity during individual transit events.  We displayed heterodyne transit signals in
two complementary formats, one of which highlights the large bandwidth and nearly quantum-noise
limited signal-to-noise ratio achieved in tracking the time-evolution of the amplitude and
phase quadratures (${\tilde x}_a,{\tilde x}_p$) of light transmitted through the cavity.  The
second data format (transit phasors) displays the correlation induced between ${\tilde x}_a$
and ${\tilde x}_p$ by the atom-cavity interaction, and our experimental results show that we
are able to distinguish unambiguously between quantum and semiclassical models of cavity QED
in only a few (or even just one) atomic transits.  We have furthermore presented the results
of rudimentary numerical simulations of atomic motion under the influence of mechanical forces
and momentum diffusion associated with the strong atom-cavity coupling, and we examined the
interpretations suggested by these simulations for certain qualitative features of our
experimental data ({\it i.e.} steps and asymmetries).

The primary conclusion we wish to draw in comparing the simulation results with our
experimental data is that we have experimentally reached a regime of measurement sensitivity
and bandwidth in which details of the atomic center-of-mass trajectories really should be
visible.  Even though we have no incontrovertible means of proving that steps and asymmetries
seen in our experimental data should be associated with dynamical processes like channeling
and diffusive escape, we are now motivated in our continuing work to develop some means of
{\em actively influencing} the atomic motion {\em while} it is still inside the cavity.  This
would allow us to produce deliberate displacements of an atom along the cavity standing wave,
and to examine the induced variations of the heterodyne signal in order to verify our inferred
displacement sensitivity of $1.5\times 10^{-10}$ m/$\sqrt{\rm Hz}$.  We anticipate that the
combined abilities of monitoring and influencing atomic position relative to the cavity
standing wave will enable the investigation of schemes for {\em real-time feedback control}
of quantized atomic center-of-mass motion.

Finally, let us note that the standard quantum limit for overall observation time of the
position of a single atom with sensitivity $1.5\times 10^{-10}$ m/$\sqrt{\rm Hz}$ and 300 kHz
bandwidth should be $t_*\sim 10$ $\mu$s \cite{Mabu98a}.  This implies that quantitative
experimental investigations of conditional quantum dynamics (as described in \cite{Mabu98a})
should indeed become possible in our cavity QED system once we are able to reliably prepare
individual atoms in well-defined initial states of motion.  The ideal situation in this regard
would be to trap and localize atoms {\em within} the cavity, releasing them at a node or
antinode of the standing wave, on the cavity axis, and with an initial position uncertainty
that is small compared to $\lambda/4$.  Current efforts in our group focus on trying to achieve
this level of control via optical dipole-force traps and/or far-detuned optical lattices inside
the cavity.

We wish to acknowledge Q. A. Turchette's vital participation in the early stages of this work,
and to thank A.~C. Doherty, C. J. Hood, and T. W. Lynn for valuable discussions.  Much of the
numerical work described in this paper was performed on resources located at the Advanced
Computing Laboratory of the Los Alamos National Laboratory.  This research was supported by
the National Science Foundation under Grant No. PHY97-22674, by the ONR, and by DARPA through
the QUIC initiative (administered by ARO).  Jun Ye is supported by an R. A. Millikan Fellowship
from the California Institute of Technology.

\begin{figure}
\caption{Schematic overview of the apparatus.}
\label{fig:apparat}
\end{figure}

\begin{figure}
\caption{Geometrical arrangement of the MOT beams relative to the physics cavity (figure not
to scale).  The mirror substrates are each 3 mm in diameter and 4 mm long.  The MOT forms at a
height of $\sim 7$ mm above the cavity axis.}
\label{fig:motconfig}
\end{figure}

\begin{figure}
\caption{A 15 ms segment of the typical recorded data, showing the simultaneously-recorded
photocurrents proportional to the amplitude quadrature ${\tilde x}_a$ (upper trace) and phase
quadrature ${\tilde x}_p$ (lower trace) of light transmitted through the physics cavity (see
text).  The experimental parameters for this data were $\Delta=10$ MHz, $\Theta=0$, and
$m=1.5$.  The photocurrents were digitized at a rate of 10 MHz, with 12-bit
resolution, and the analog bandwidth of the anti-aliasing filters was 300 kHz.  Note the set of
transient features clustered between $0.009$ and $0.013$ on the time axis, each of which was
caused by the passage of an individual atom through the cavity.}
\label{fig:hmtlong}
\end{figure}

\begin{figure}
\caption{Cavity diffusion coefficients versus atomic position, for (from the top curve going
down) atom-probe detuning $\Delta\equiv\nu_a-\nu_p=0$, $10$, $30$, and $50$ MHz, with $m=2$
photons and $\Theta=0$.}
\label{fig:cavdiff1}
\end{figure}

\begin{figure}
\caption{Cavity diffusion coefficients versus atomic position, for (from the top curve going
down) $m=8$, $4$, and $2$ photons, with the atom-probe detuning $\Delta=50$
MHz (again $\Theta=0$).}
\label{fig:cavdiff2}
\end{figure}

\begin{figure}
\caption{(a) Simulated atomic trajectory and corresponding heterodyne signal for a single
transit with $m=4$, $\Theta=0$, and $\Delta=50$ MHz.  Here $x$ is atomic position along the
cavity standing-wave (measured in units of the optical wavelength) and $v_x$ is the atomic
velocity along the standing-wave (measured in optical wavelengths per $\mu$s).  Note that the
simulation includes three-dimensional classical center-of-mass motion for the atom, although
only $x$ and $v_x$ are displayed above.  The two subplots in part (b) display the expectation
value of the intracavity phase quadrature amplitude $\langle q_p\rangle$ (which is proportional
to ${\tilde x}_p$) filtered down to an analog bandwidth $\sim 300$ kHz (left), as well as
$\langle q_p\rangle$ plus an appropriate amount of Gaussian noise to simulate shot noise in our
heterodyne detection (right).}
\label{fig:simex1}
\end{figure}

\begin{figure}
\caption{Simulated atom-transit signals (see text), displayed at an analog bandwidth of 200
kHz.  The atom-probe detuning $\Delta$ and probe power $m$ are indicated for each subplot, and
the atom-cavity detuning $\Theta$ is zero in all cases.  Upper traces represent the intracavity
amplitude quadrature $\langle q_a\rangle$, with Gaussian noise added to reproduce the overall
signal-to-noise ratio predicted by equation (10) (yielding a quantity proportional to
${\tilde x}_a$).  Lower traces (with zero mean) represent the intracavity phase quadrature
$\langle q_p\rangle$ plus Gaussian noise (proportional to ${\tilde x}_p$).}
\label{fig:sims}
\end{figure}

\begin{figure}
\caption{Individual atom-transit signals, displayed at the full analog bandwidth of 300 kHz and
sampling rate of 10 MHz (12 bit resolution).  The atom-probe detuning $\Delta$ and probe power
$m$ (see text) are indicated for each subplot, and the cavity-probe detuning $\Theta$ is zero
in all cases.  Upper traces represent the amplitude quadrature ${\tilde x}_a$, whereas the
lower traces (with zero mean) represent the phase quadrature ${\tilde x}_p$.  Note that we
have displaced the ${\tilde x}_a$ traces by $+400$ in order to prevent them from overlapping
with the ${\tilde x}_p$ traces, and that the photocurrents are displayed in arbitrary units.}
\label{fig:hmthbw}
\end{figure} 

\begin{figure}
\caption{Dependence of transit phasor shapes on detuning (the values of probe power $m$ and
detuning $\Delta$ are indicated above each subplot).  Each subplot displays an overlay of
two data traces (gray spots), quantum-mechanical theory from the Master Equation (solid curve
ending in a $\circ$, computed using equation (1)), and semiclassical theory (computed
using equation (7), solid curve ending in an ``$\times$'').}
\label{fig:hmtfig3}
\end{figure}

\begin{figure}
\caption{Transit phasors for fixed detuning $\Delta=10$ MHz, with variable probe strength
(as indicated).  Each subplot displays an overlay of three data traces (gray spots),
quantum-mechanical theory from the Master Equation (solid curve ending in a $\circ$, computed
using equation (1)), and semiclassical theory (computed using equation
(7), solid curve ending in an ``$\times$'').}
\label{fig:hmtfig2}
\end{figure}

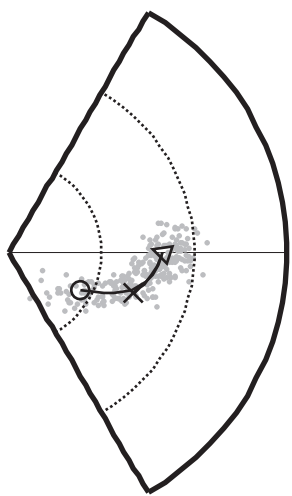
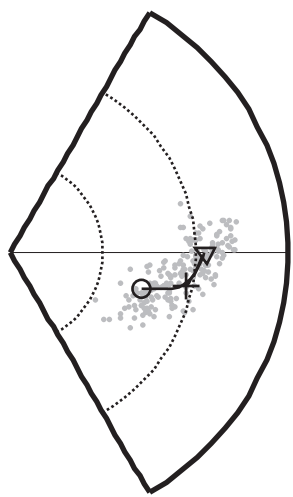
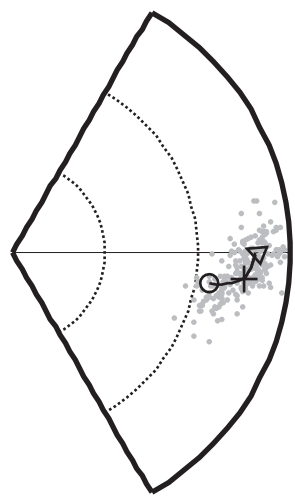
\end{document}